\documentclass[a4paper]{jpconf}

\usepackage[centertags]{amsmath}
\allowdisplaybreaks[1]
\usepackage{amsbsy}
\usepackage{amsfonts}
\usepackage{amssymb}
\usepackage[dvips]{graphicx}
\usepackage{cite}

\newcommand{\boss}[2]{\ensuremath{\rlap{\kern-2.5pt\ensuremath{\overset{\scriptscriptstyle(-)}{\phantom{#1}}}}{\ensuremath{{#1}_{#2}}}}}

\begin{document}

\title{Phenomenology of Sterile Neutrinos}

\author{Carlo Giunti}

\address{INFN, Sezione di Torino, Via P. Giuria 1, I--10125 Torino, Italy}

\ead{giunti@to.infn.it}

\begin{abstract}
The indications
in favor of short-baseline neutrino oscillations,
which require the existence of one or more sterile neutrinos,
are reviewed.
In the framework of 3+1 neutrino mixing,
which is the simplest extension of the standard three-neutrino mixing which can partially explain the data,
there is a strong tension in the interpretation of the data,
mainly due to an incompatibility of the results of appearance and disappearance experiments.
In the framework of 3+2 neutrino mixing,
CP violation in short-baseline experiments
can explain the difference between
MiniBooNE neutrino and antineutrino data,
but
the tension
between the data of appearance and disappearance experiments persists
because the short-baseline
disappearance of electron antineutrinos and muon neutrinos
compatible with the LSND and MiniBooNE antineutrino appearance signal has not been observed.
\\
\textit{
Invited paper to
NUFACT 11, XIIIth International Workshop on Neutrino Factories, Super beams and Beta beams, 1-6 August 2011, CERN and University of Geneva
(Submitted to IOP conference series).}
\end{abstract}

\section{Introduction}
\label{Introduction}

From the results of solar, atmospheric and long-baseline
neutrino oscillation experiments we know that
neutrinos are massive and mixed particles
(see Ref.~\cite{Giunti-Kim-2007}).
There are two groups of experiments which measured
two independent squared-mass differences
($\Delta{m}^2$)
in two different neutrino flavor transition channels:

\begin{itemize}

\item
Solar neutrino experiments
(Homestake,
Kamiokande,
GALLEX/GNO,
SAGE,
Super-Kamiokande,
SNO,
BOREXino)
measured $\nu_{e} \to \nu_{\mu}, \nu_{\tau}$
oscillations generated by
$
\Delta m^2_{\text{SOL}}
=
6.2 {}^{+1.1}_{-1.9} \times 10^{-5} \, \text{eV}^2
$
and a mixing angle
$
\tan^2 \vartheta_{\text{SOL}}
=
0.42 {}^{+0.04}_{-0.02}
$
\cite{1010.0118}.
The KamLAND experiment
confirmed these oscillations by observing the disappearance
of reactor $\bar\nu_{e}$ at an average distance of about 180 km.
The combined fit of solar and KamLAND data leads to
$
\Delta m^2_{\text{SOL}}
=
(7.6 \pm 0.2) \times 10^{-5} \, \text{eV}^2
$
and a mixing angle
$
\tan^2 \vartheta_{\text{SOL}}
=
0.44 \pm 0.03
$
\cite{1010.0118}.

\item
Atmospheric neutrino experiments
(Kamiokande,
IMB,
Super-Kamiokande,
MACRO,
Soudan-2,
MINOS)
measured $\nu_{\mu}$ and $\bar\nu_{\mu}$
disappearance through oscillations generated by
$
\Delta m^2_{\text{ATM}}
\simeq
2.3 \times 10^{-3} \, \text{eV}^2
$
and a mixing angle
$
\sin^2 2\vartheta_{\text{ATM}}
\simeq
1
$
\cite{hep-ex/0501064}.
The K2K and MINOS long-baseline experiments
confirmed these oscillations by observing the disappearance
of accelerator $\nu_{\mu}$
at distances of about 250 km and 730 km, respectively.
The MINOS data give
$
\Delta m^2_{\text{ATM}}
=
2.32 {}^{+0.12}_{-0.08} \times 10^{-3} \, \text{eV}^2
$
and
$
\sin^2 2\vartheta_{\text{ATM}}
>
0.90
$
at 90\% C.L.
\cite{1103.0340}.

\end{itemize}

These measurements led to the current
three-neutrino mixing paradigm,
in which the three active neutrinos
$\nu_{e}$,
$\nu_{\mu}$,
$\nu_{\tau}$
are superpositions of three massive neutrinos
$\nu_1$,
$\nu_2$,
$\nu_3$
with respective masses
$m_1$,
$m_2$,
$m_3$.
The two measured squared-mass differences can be interpreted as
\begin{equation}
\Delta m^2_{\text{SOL}}
=
\Delta m^2_{21}
\,,
\qquad
\Delta m^2_{\text{ATM}}
=
|\Delta m^2_{31}|
\simeq
|\Delta m^2_{32}|
\,,
\label{dm2}
\end{equation}
with
$\Delta m^2_{kj}=m_k^2-m_j^2$.
In the standard parameterization of the $3\times3$ unitary mixing matrix
(see Ref.~\cite{Giunti-Kim-2007})
$\vartheta_{\text{SOL}} \simeq \vartheta_{12}$,
$\vartheta_{\text{ATM}} \simeq \vartheta_{23}$
and
$\sin^2\vartheta_{13}<0.035$
at 90\% C.L.
\cite{0808.2016}.

\begin{figure}[t!]
\begin{center}
\includegraphics*[bb=76 546 571 767, width=0.6\textwidth]{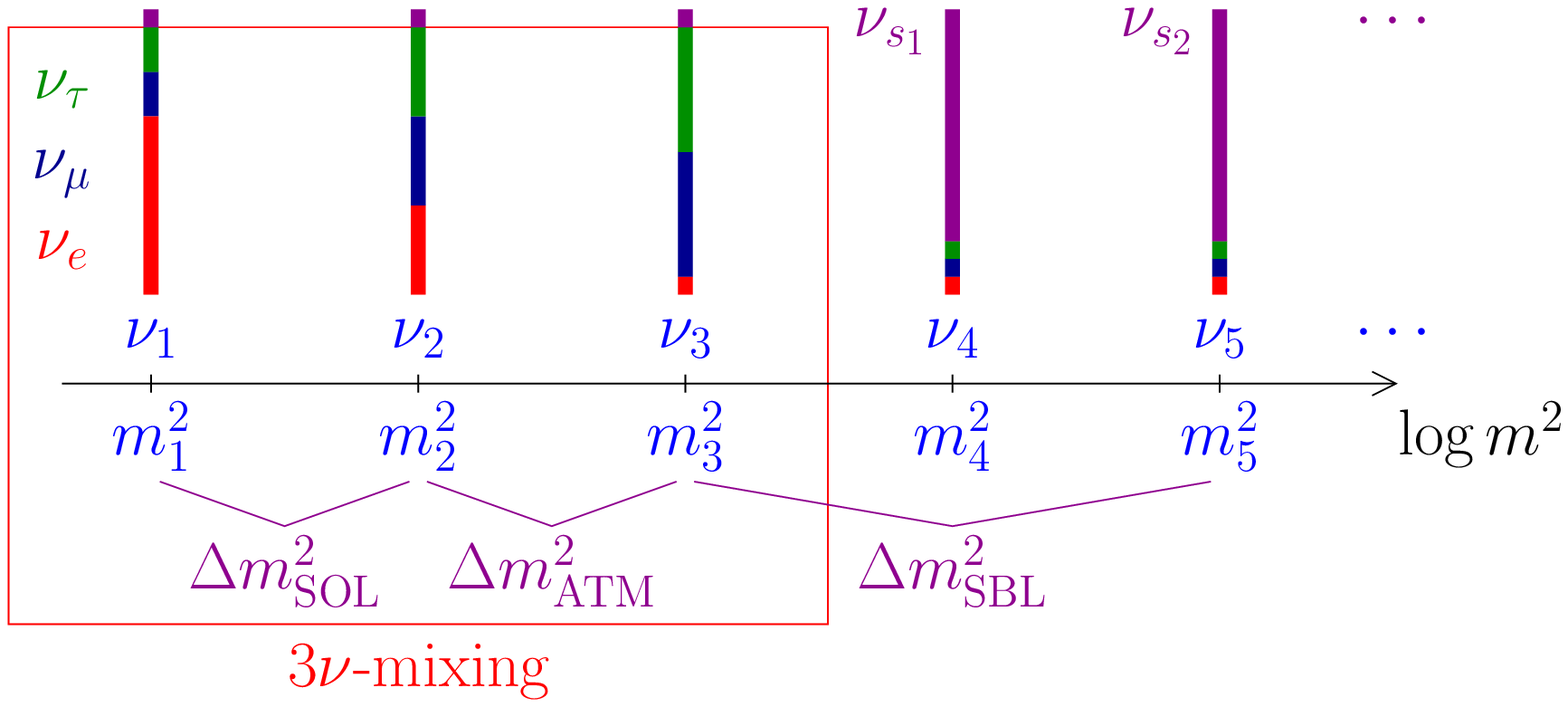}
\null
\vspace{-0.5cm}
\null
\end{center}
\caption{\small \label{nuspec}
Schematic description of the neutrino spectrum beyond three-neutrino mixing with one or more relatively heavy additional massive neutrinos
which are mainly sterile.
}
\end{figure}

The completeness of the three-neutrino mixing paradigm was
challenged in 1995 by the observation of
a signal of short-baseline
$\bar\nu_{\mu}\to\bar\nu_{e}$
oscillations in the
LSND experiment
\cite{nucl-ex/9504002,hep-ex/0104049},
which would imply the existence of
one or more squared-mass differences much larger than
$\Delta m^2_{\text{SOL}}$
and
$\Delta m^2_{\text{ATM}}$.
The MiniBooNE experiment was made in order to check the LSND signal
with about one order of magnitude larger distance ($L$) and energy ($E$),
but the same order of magnitude for the ratio $L/E$
from which neutrino oscillations depend.
The first results of the MiniBooNE experiment
in neutrino mode did not show a signal compatible with that of LSND
\cite{0812.2243},
but the results in antineutrino mode, presented in the summer of 2010 \cite{1007.1150},
show an excess of events over the background
at approximately at the same $L/E$ of LSND.
This result revived the interest
in the possibility of existence of
one or more neutrinos with masses at the eV scale
which can generate squared-mass differences for short-baseline oscillations.

Figure~\ref{nuspec} illustrates schematically the neutrino spectrum beyond three-neutrino mixing with one or more additional massive neutrinos
heavier than the three standard massive neutrinos.
In the flavor basis the additional massive neutrinos correspond to
sterile neutrinos,
which do not have standard weak interactions and do not contribute
to the number of active neutrinos determined by LEP experiments
through the measurement of the invisible width of the $Z$ boson,
$N_{a} = 2.9840 \pm 0.0082$
\cite{hep-ex/0509008}.
The existence of sterile neutrinos
which have been thermalized in the early Universe is compatible
with Big-Bang Nucleosynthesis data
\cite{astro-ph/0408033,1001.4440},
with the indication however that schemes with more than one sterile neutrino are disfavored
\cite{1103.1261}.
It is also compatible
with cosmological measurements of the
Cosmic Microwave Background and Large-Scale Structures
if the neutrino masses are limited below about 1 eV
\cite{1006.5276,1102.4774,1104.0704,1104.2333,1106.5052}.
Therefore,
we consider neutrino mixing schemes in which the three standard neutrinos have masses much smaller that 1 eV
and the additional neutrinos have masses at the eV scale.
However,
other sterile neutrinos may exist and the cosmological constraints can be avoided
by suppressing the thermalization of sterile neutrinos in the early Universe
and/or
by considering non-standard cosmological theories.
For example,
a sterile neutrino with a mass scale much smaller than 1 eV
could have important implications for solar neutrino oscillations
(see Ref.~\cite{1012.5627})
and a sterile neutrino with a mass at the keV scale could constitute Warm Dark Matter
(see Refs.~\cite{0906.2968,0901.0011}).

A further indication in favor of short-baseline oscillations
came from a new calculation of the reactor $\bar\nu_{e}$ flux presented in January 2011 \cite{1101.2663}
(see also Ref.~\cite{1106.0687}),
which obtained an increase of about 3\% with respect to the previous value adopted
in the analysis of the data of reactor neutrino oscillation experiments.
As illustrated in Fig.~2 of Ref.~\cite{1106.4479},
the measured reactor rates are in agreement with those derived from the old $\bar\nu_{e}$ flux,
but show a deficit of about $2.2\sigma$ with respect to the rates derived from the new $\bar\nu_{e}$ flux.
This is the ``reactor antineutrino anomaly'' \cite{1101.2755},
which may be an indication in the
$\bar\nu_{e}\to\bar\nu_{e}$
channel of a signal corresponding to the
$\bar\nu_{\mu}\to\bar\nu_{e}$
signal observed in the
LSND and MiniBooNE experiments.
Finally,
there is a ``Gallium neutrino anomaly''
\cite{hep-ph/0610352,0707.4593,0711.4222,0902.1992,1005.4599,1006.3244},
consisting in a short-baseline disappearance of electron neutrinos
measured in the
Gallium radioactive source experiments
GALLEX
\cite{1001.2731}
and
SAGE
\cite{0901.2200}.

In the following,
I consider in Sections~\ref{3+1}
and
\ref{3+2}
the cases of 3+1
\cite{hep-ph/9606411,hep-ph/9607372,hep-ph/9903454,hep-ph/0405172}
and
3+2
\cite{hep-ph/0305255,hep-ph/0609177,0705.0107,0906.1997}
neutrino mixing, respectively,
following the discussion in Ref.~\cite{1107.1452}.
Conclusions are drawn in Section~\ref{Conclusions}.

\section{3+1 Neutrino Mixing}
\label{3+1}

\begin{table}[t!]
\begin{center}
\small
\begin{tabular}{ccc}
&
3+1
&
3+2
\\
\hline
$\chi^2_{\text{min}}$
&
$100.2$
&
$91.6$
\\
$\text{NDF}$
&
$104$
&
$100$
\\
$\text{GoF}$
&
$59\%$
&
$71\%$
\\
\hline
$\Delta{m}^2_{41}\,[\text{eV}^2]$
&
$0.89$
&
$0.90$
\\
$|U_{e4}|^2$
&
$0.025$
&
$0.017$
\\
$
|U_{\mu4}|^2$
&
$0.023$
&
$0.019$
\\
$\Delta{m}^2_{51}\,[\text{eV}^2]$
&
&
$1.61$
\\
$|U_{e5}|^2$
&
&
$0.017$
\\
$
|U_{\mu5}|^2$
&
&
$0.0061$
\\
$\eta$
&
&
$1.51 \pi$
\\
\hline
$\Delta\chi^{2}_{\text{PG}}$
&
$24.1$
&
$22.2$
\\
$\text{NDF}_{\text{PG}}$
&
$2$
&
$5$
\\
$\text{PGoF}$
&
$
6
\times
10^{-6}
$
&
$
5
\times
10^{-4}
$
\\
\hline
\end{tabular}
\null
\vspace{-0.3cm}
\null
\end{center}
\caption{\small \label{bef}
Values of
$\chi^{2}$,
number of degrees of freedom (NDF),
goodness-of-fit (GoF)
and
best-fit values of the mixing parameters
obtained in our 3+1 and 3+2 fits of short-baseline oscillation data.
The last three lines give the results of the parameter goodness-of-fit test
\cite{hep-ph/0304176}:
$\Delta\chi^{2}_{\text{PG}}$,
number of degrees of freedom ($\text{NDF}_{\text{PG}}$) and
parameter goodness-of-fit (PGoF).
}
\end{table}

In this Section I consider the
simplest extension of three-neutrino mixing
with the addition of one massive neutrino.
In 3+1 neutrino mixing
\cite{hep-ph/9606411,hep-ph/9607372,hep-ph/9903454,hep-ph/0405172}
the
effective flavor transition and survival probabilities
in short-baseline (SBL) experiments
are given by
\begin{equation}
P_{\boss{\nu}{\alpha}\to\boss{\nu}{\beta}}^{\text{SBL}}
=
\sin^{2} 2\vartheta_{\alpha\beta}
\sin^{2}\left( \frac{\Delta{m}^{2}_{41} L}{4E} \right)
\,,
\qquad
P_{\boss{\nu}{\alpha}\to\boss{\nu}{\alpha}}^{\text{SBL}}
=
1
-
\sin^{2} 2\vartheta_{\alpha\alpha}
\sin^{2}\left( \frac{\Delta{m}^{2}_{41} L}{4E} \right)
\,,
\label{3p1-pro}
\end{equation}
for
$\alpha,\beta=e,\mu,\tau,s$
and
$\alpha\neq\beta$,
with
\begin{equation}
\sin^{2} 2\vartheta_{\alpha\beta}
=
4 |U_{\alpha4}|^2 |U_{\beta4}|^2
\,,
\qquad
\sin^{2} 2\vartheta_{\alpha\alpha}
=
4 |U_{\alpha4}|^2 \left( 1 - |U_{\alpha4}|^2 \right)
\,.
\label{3p1-amp}
\end{equation}
Therefore:
\begin{enumerate}
\renewcommand{\theenumi}{\arabic{enumi}}
\item
All effective SBL oscillation probabilities
depend only on the absolute value of the largest squared-mass difference
$\Delta{m}^2_{41}=m_{4}^2-m_{1}^2$.
\item
All oscillation channels are open, each one with its own oscillation amplitude.
\item
The oscillation amplitudes depend only on the absolute values
of the elements in the fourth column of the mixing matrix,
i.e. on three real numbers with sum less than unity,
since the unitarity of the mixing matrix implies that
$
\sum_{\alpha} |U_{\alpha4}|^2 = 1
$.
\item
CP violation cannot be observed in SBL oscillation experiments,
even if the mixing matrix contains CP-violation phases,
because
neutrinos and antineutrinos have the same
effective SBL oscillation probabilities.
Hence,
3+1 neutrino mixing
cannot explain the difference between
neutrino \cite{0812.2243} and antineutrino \cite{1007.1150} oscillations observed in the
MiniBooNE.
\end{enumerate}

\begin{figure}[t!]
\null
\hfill
\includegraphics*[width=0.45\textwidth]{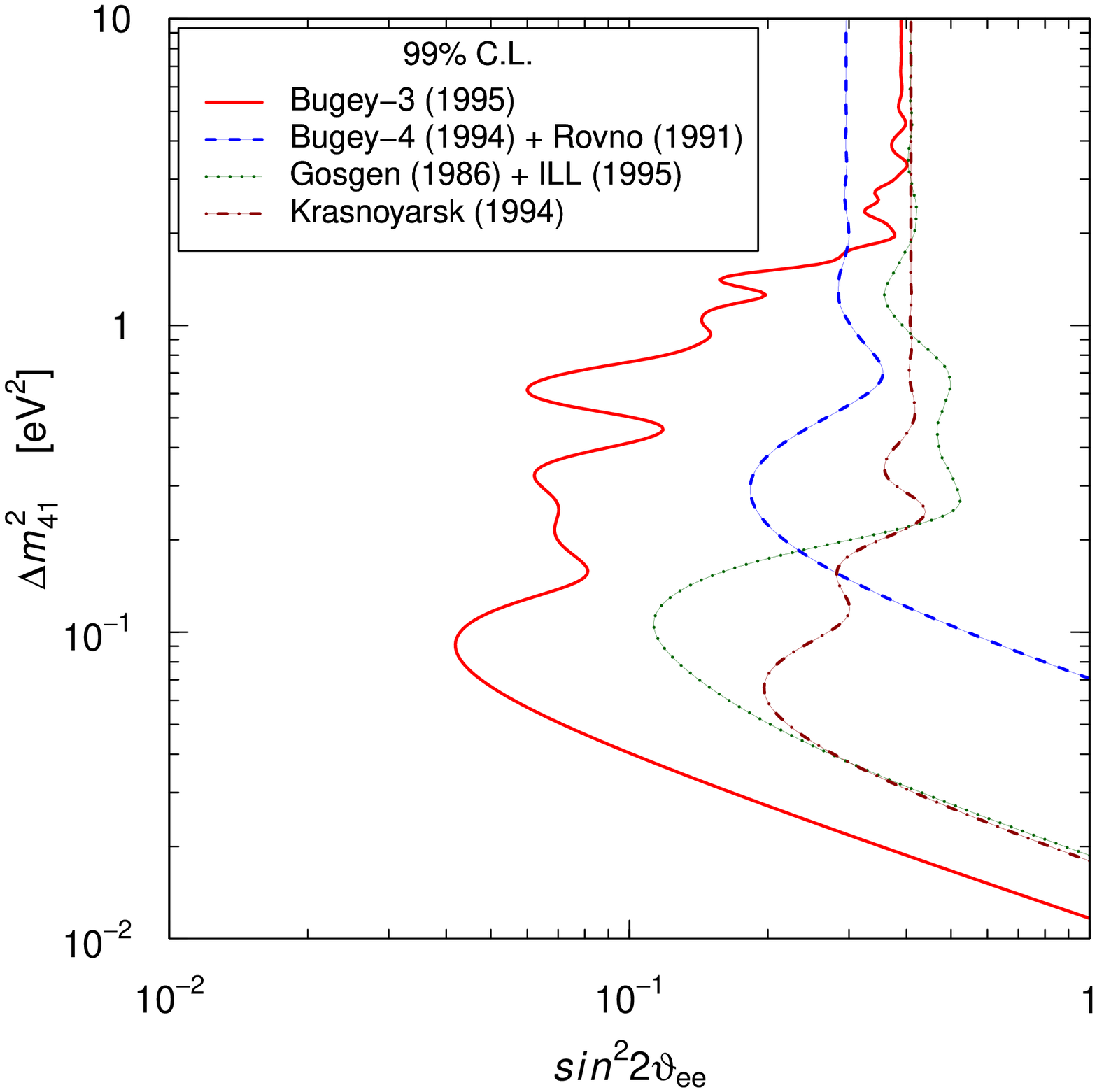}
\hfill
\includegraphics*[width=0.45\textwidth]{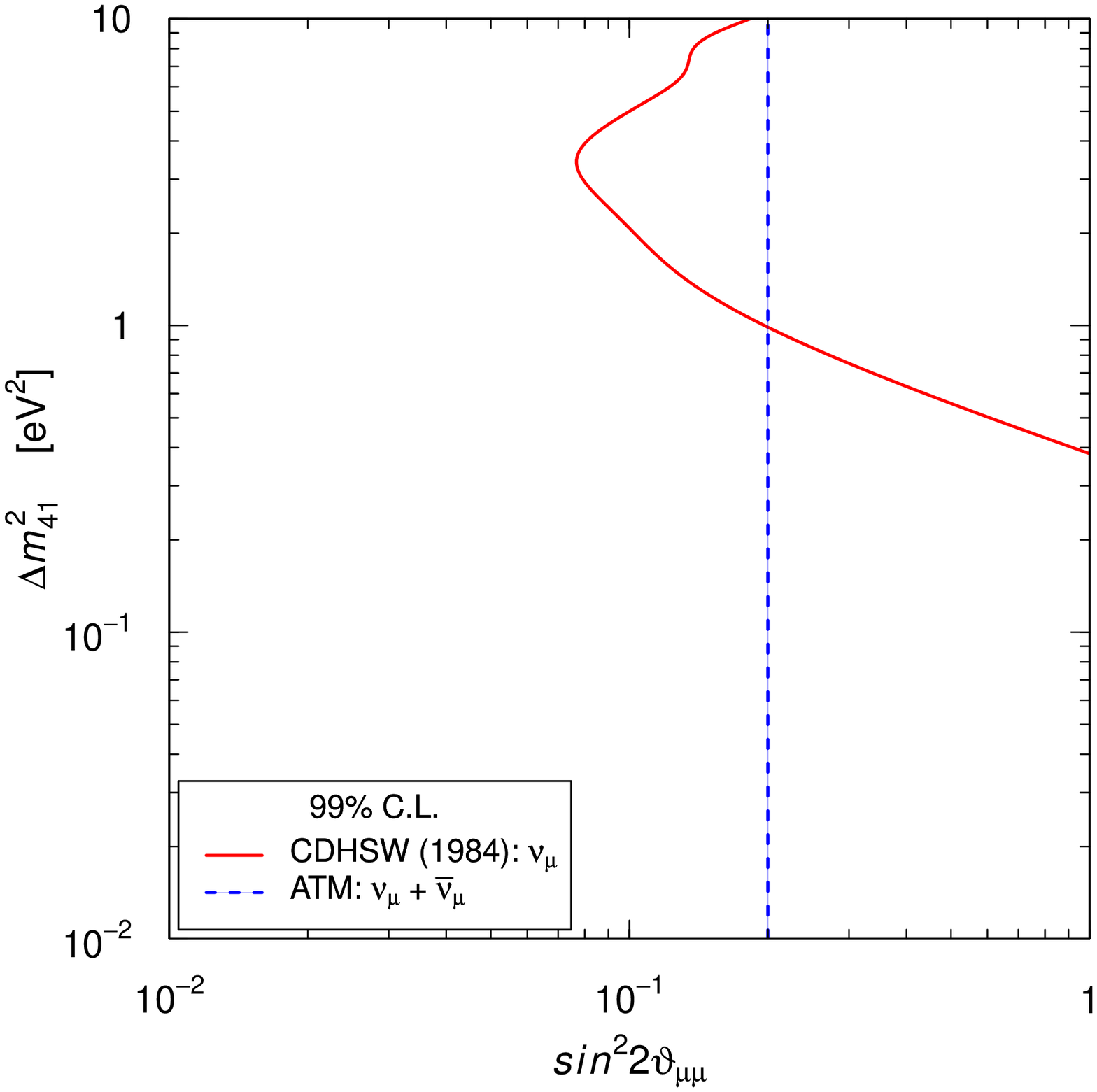}
\hfill
\null
\vspace{-0.3cm}
\null
\caption{\small \label{dis}
Exclusion curves obtained from the data of reactor $\bar\nu_{e}$ disappearance experiments
(see Ref.~\cite{1101.2755}),
from the data of the CDHSW $\nu_{\mu}$ disappearance experiment
\cite{Dydak:1984zq},
and from atmospheric neutrino data
(extracted from the analysis in Ref.~\cite{0705.0107}).
}
\end{figure}

The dependence of the oscillation amplitudes in Eq.~(\ref{3p1-amp})
on three independent absolute values
of the elements in the fourth column of the mixing matrix
implies that the amplitude of
$
\boss{\nu}{\mu}\to\boss{\nu}{e}
$
transitions is limited by the absence of large SBL disappearance of
$\boss{\nu}{e}$ and $\boss{\nu}{\mu}$
observed in several experiments.

The results of reactor neutrino experiments constrain the value
$|U_{e4}|^2$
through the measurement of
$\sin^{2} 2\vartheta_{ee}$.
Even taking into account the reactor antineutrino anomaly \cite{1101.2755}
discussed in the Introduction,
the $\bar\nu_{e}$ disappearance is small and large values of
$\sin^{2} 2\vartheta_{ee}$
are constrained by the exclusion curves in the left panel of Fig.~\ref{dis}.
Since values of $|U_{e4}|^2$ close to unity are excluded by solar neutrino oscillations
(which require large $|U_{e1}|^2+|U_{e2}|^2$),
for small $\sin^{2} 2\vartheta_{ee}$ we have
\begin{equation}
\sin^{2} 2\vartheta_{ee} \simeq 4 |U_{e4}|^2
\,.
\label{ue4}
\end{equation}

The value of $\sin^{2} 2\vartheta_{\mu\mu}$
is constrained by the curves in the right panel of Fig.~\ref{dis},
which have been obtained from
the lack of $\nu_{\mu}$ disappearance in the CDHSW $\nu_{\mu}$ experiment
\cite{Dydak:1984zq}
and
from the requirement of large $|U_{\mu1}|^2+|U_{\mu2}|^2+|U_{\mu3}|^2$
for atmospheric neutrino oscillations \cite{0705.0107}.
Hence,
$|U_{\mu4}|^2$ is small and
\begin{equation}
\sin^{2} 2\vartheta_{\mu\mu} \simeq 4 |U_{\mu4}|^2
\,.
\label{um4}
\end{equation}

\begin{figure}[t!]
\null
\hfill
\includegraphics*[width=0.45\textwidth]{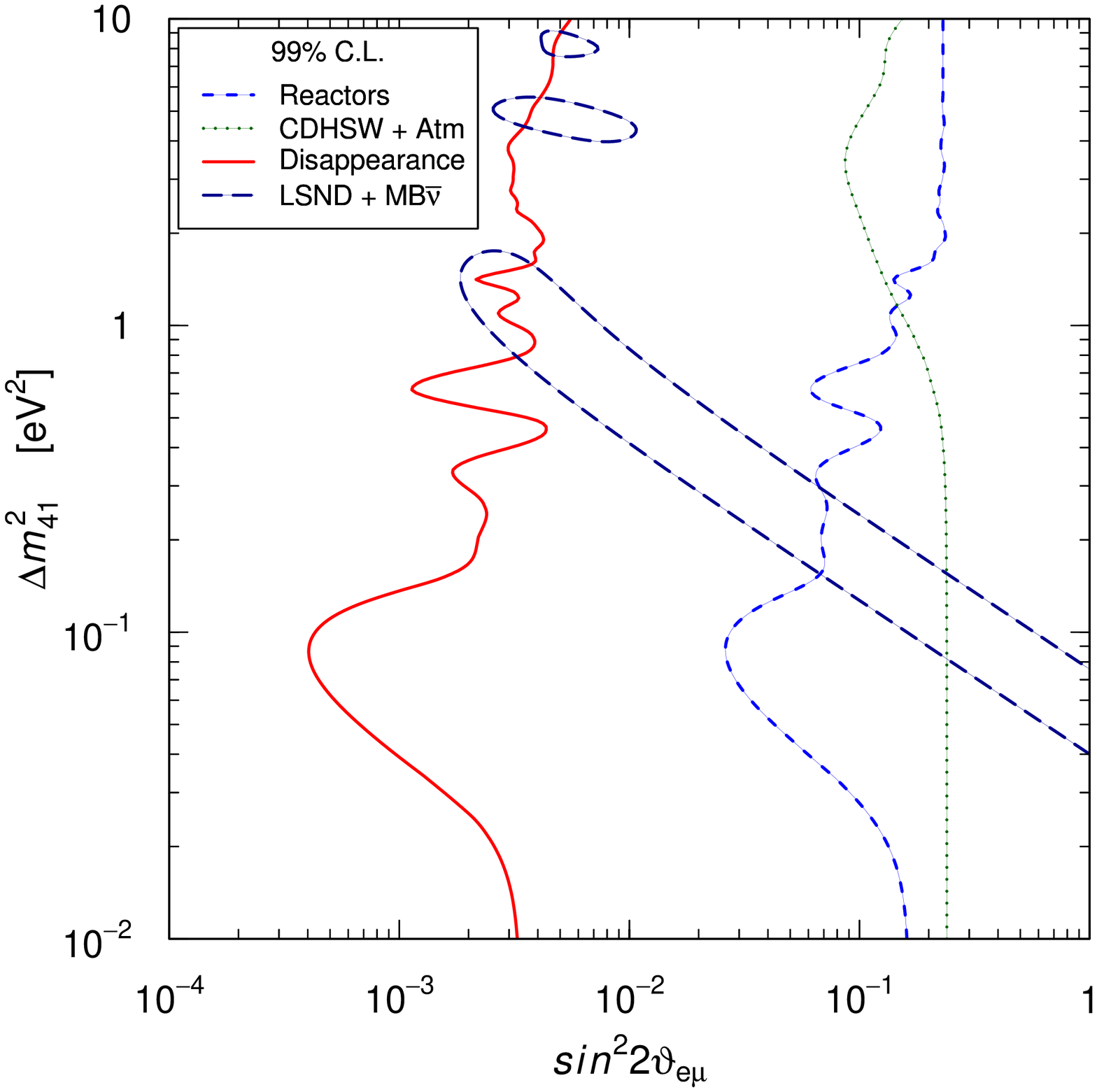}
\hfill
\includegraphics*[width=0.45\textwidth]{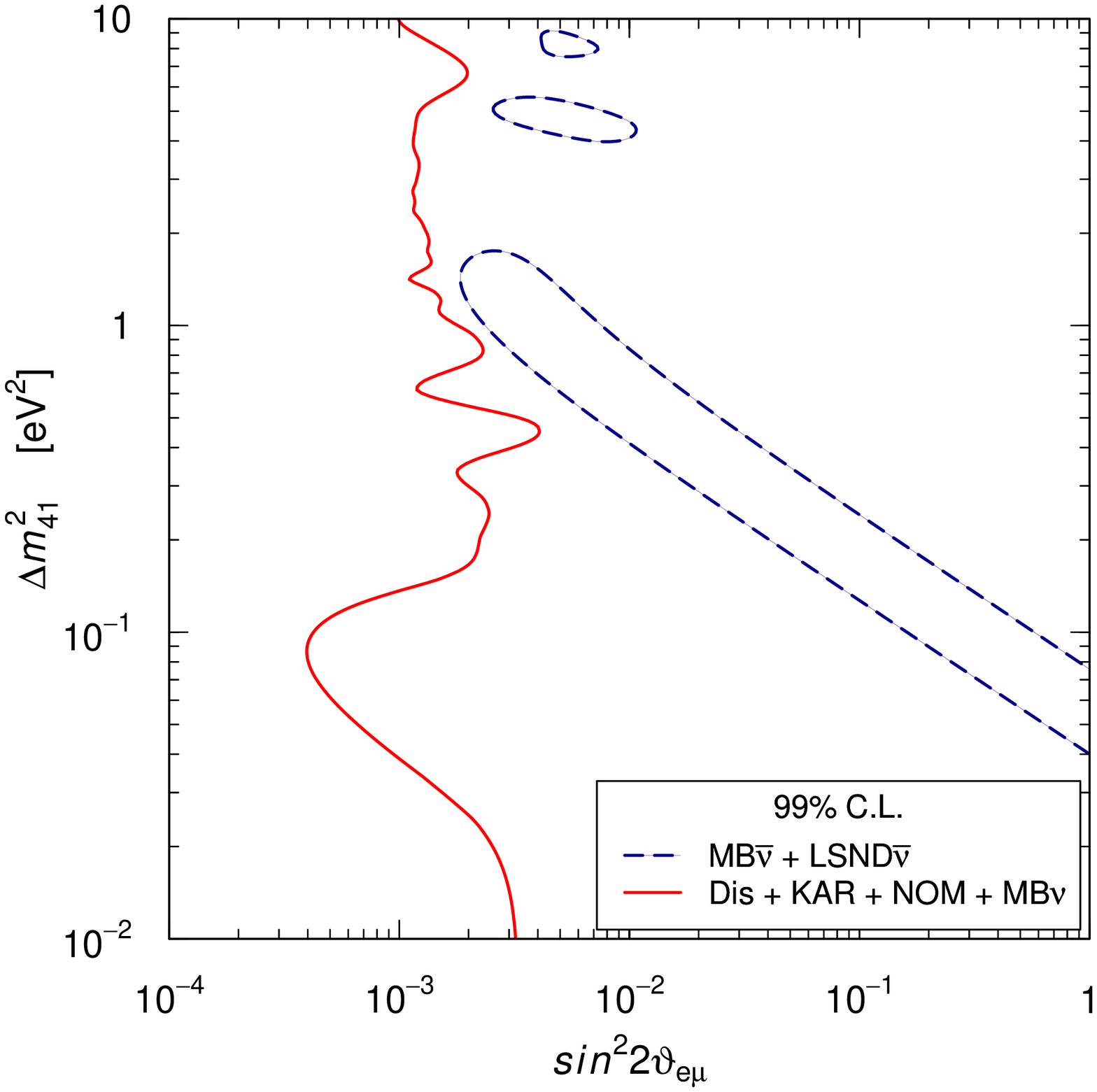}
\hfill
\null
\vspace{-0.3cm}
\null
\caption{\small \label{exc}
Left panel:
Exclusion curves
in the $\sin^{2} 2\vartheta_{e\mu}$--$\Delta{m}^2_{41}$ plane
obtained from the separate constraints in Fig.~\ref{dis}
(blue and green lines)
and the combined constraint given by Eq.~(\ref{sem})
(red line)
from disappearance experiments (Dis).
Right panel:
Exclusion curve obtained with the addition of
KARMEN \cite{hep-ex/0203021} (KAR),
NOMAD \cite{hep-ex/0306037} (NOM)
and 
MiniBooNE neutrino \cite{0812.2243} (MB$\nu$)
data (red line).
In both panels the region enclosed by the dark-red lines
is allowed by
LSND and MiniBooNE antineutrino data.
}
\end{figure}

From Eqs.~(\ref{3p1-amp}), (\ref{ue4}) and (\ref{um4}),
for the amplitude of
$
\boss{\nu}{\mu}\to\boss{\nu}{e}
$
transitions we obtain
\begin{equation}
\sin^{2} 2\vartheta_{e\mu}
\simeq
\frac{1}{4}
\,
\sin^{2} 2\vartheta_{ee}
\,
\sin^{2} 2\vartheta_{\mu\mu}
\,.
\label{sem}
\end{equation}
Therefore,
if
$\sin^{2} 2\vartheta_{ee}$
and
$\sin^{2} 2\vartheta_{\mu\mu}$
are small,
$\sin^{2} 2\vartheta_{e\mu}$
is quadratically suppressed
\cite{hep-ph/9606411,hep-ph/9607372}.
This is illustrated in the left panel of Fig.~\ref{exc},
where one can see that the separate effects of the constraints on
$\sin^{2} 2\vartheta_{ee}$
and
$\sin^{2} 2\vartheta_{\mu\mu}$
exclude only the large-$\sin^{2} 2\vartheta_{e\mu}$
part of the region allowed by
LSND and MiniBooNE antineutrino data,
whereas most of this region is excluded by the combined constraint in Eq.~(\ref{sem}).
As shown in the right panel of Fig.~\ref{exc},
the constraint becomes stronger by including the data of the
KARMEN \cite{hep-ex/0203021},
NOMAD \cite{hep-ex/0306037}
and 
MiniBooNE neutrino \cite{0812.2243}
experiments,
which did not observe a short-baseline
$
\boss{\nu}{\mu}\to\boss{\nu}{e}
$
signal.
Since the parameter goodness-of-fit
\cite{hep-ph/0304176}
is
$
6
\times
10^{-6}
$
\cite{1107.1452},
3+1 neutrino mixing is disfavored by the data.
This conclusion has been reached recently in Refs.~\cite{1007.4171,1012.0267,1103.4570,1107.1452}
and confirms the pre-MiniBooNE antineutrino results in Refs.~\cite{hep-ph/9606411,hep-ph/9607372,hep-ph/9903454,hep-ph/0207157,hep-ph/0405172,0705.0107,0906.1997}.

However,
in spite of the low value of the parameter goodness-of-fit
it is not inconceivable to refuse to reject 3+1 neutrino mixing
for the following reasons:

\begin{enumerate}
\renewcommand{\theenumi}{\Alph{enumi}}
\item
It is the simplest scheme beyond the standard three-neutrino mixing which can partially explain the data.
\item
It corresponds to the natural addition of one new entity (a sterile neutrino) to explain a new effect
(short-baseline oscillations).
Better fits of the data require the addition of at least another new entity
(in any case at least one sterile neutrino is needed to generate short-baseline oscillations).
\item
The minimum value of the global $\chi^2$ is rather good:
$\chi^2_{\text{min}} = 100.2$
for
$104$
degrees of freedom.
\item
There is a marginal appearance--disappearance compatibility:
$
\Delta\chi^{2}_{\text{PG}}
=
9.2
$
with
2
degrees of freedom,
corresponding to
$\text{PGoF} = 1.0\%$.
\item
3+1 mixing is favored with respect to 3+2 mixing by the
Big-Bang Nucleosynthesis limit
$N_{\text{eff}}\leq4$ at 95\% C.L.
obtained in Ref.~\cite{1103.1261}.
\end{enumerate}

Therefore,
we consider the global fit of all data in the framework of 3+1 neutrino mixing,
which yields the best-fit values of the oscillation parameters listed in Tab.~\ref{bef}.

\begin{figure}[t!]
\begin{center}
\begin{tabular}{cc}
\includegraphics*[width=0.45\linewidth]{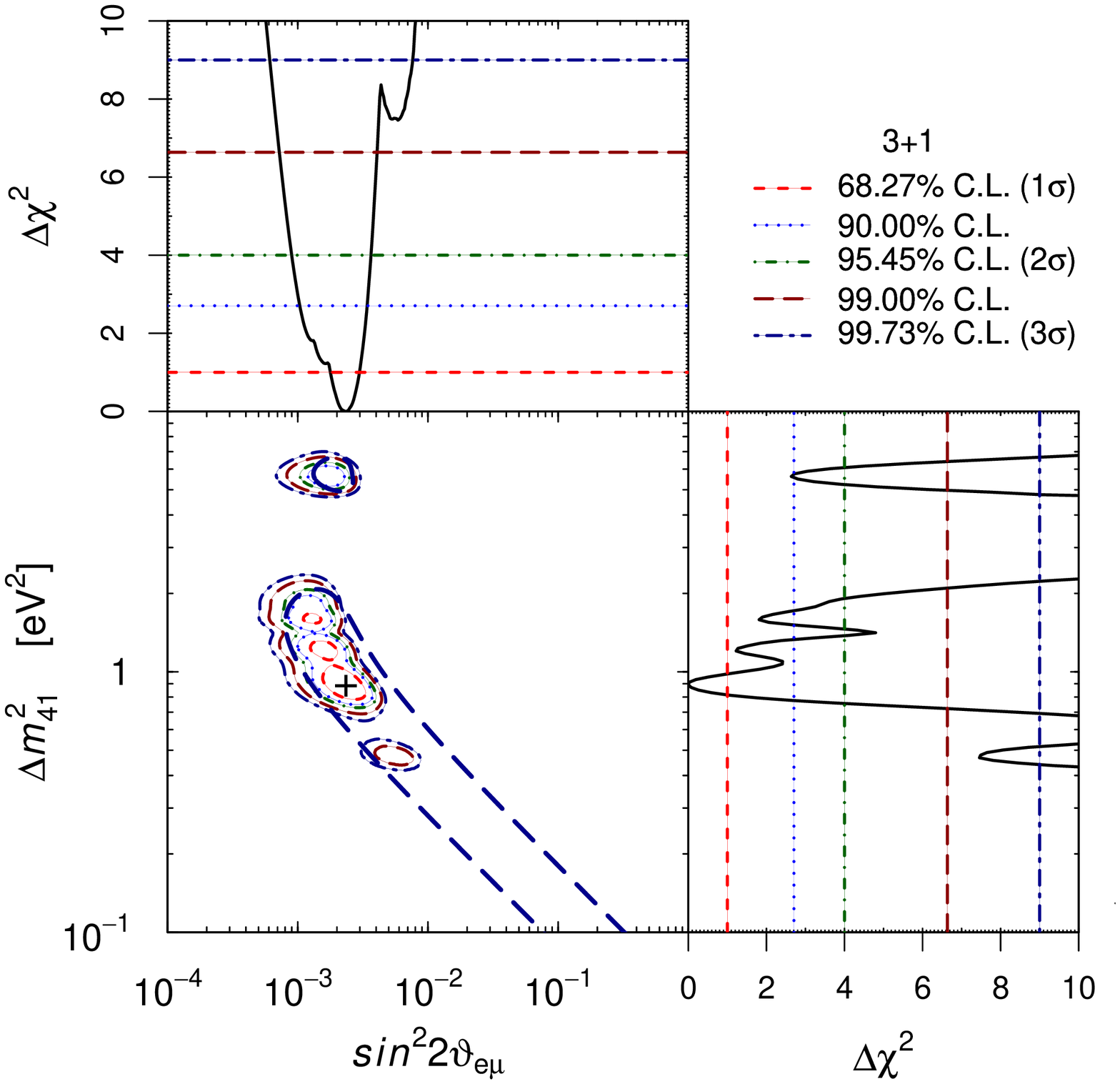}
&
\includegraphics*[width=0.45\linewidth]{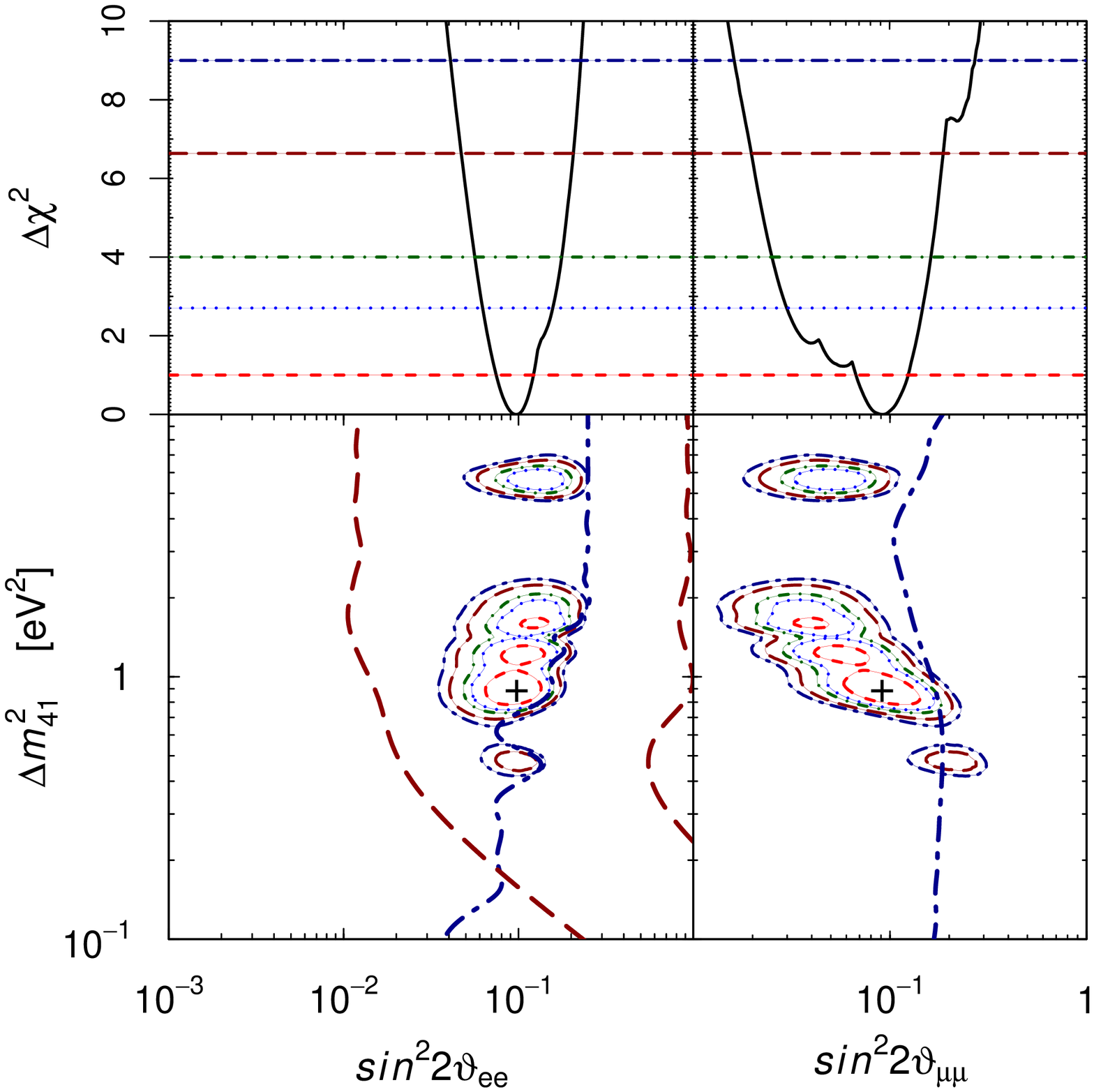}
\end{tabular}
\null
\vspace{-0.5cm}
\null
\end{center}
\caption{\small \label{3p1-fig}
Allowed regions in the
$\sin^{2}2\vartheta_{e\mu}$--$\Delta{m}^2_{41}$,
$\sin^{2}2\vartheta_{ee}$--$\Delta{m}^2_{41}$ and
$\sin^{2}2\vartheta_{\mu\mu}$--$\Delta{m}^2_{41}$ planes
and
marginal $\Delta\chi^{2}$'s
obtained from the global fit in 3+1 neutrino mixing.
The best-fit point is indicated by a cross.
Left panel:
the isolated dark-blue long-dashed contours enclose the regions allowed at $3\sigma$
by the analysis of appearance data
(LSND \cite{hep-ex/0104049},
KARMEN \cite{hep-ex/0203021},
NOMAD \cite{hep-ex/0306037},
MiniBooNE \cite{0812.2243,1007.1150}).
Right panel:
the isolated dark-blue dash-dotted lines are the $3\sigma$ exclusion curves
obtained from reactor neutrino data
and from CDHSW and atmospheric neutrino data.
The isolated dark-red long-dashed lines delimit the region allowed at 99\% C.L.
by the Gallium anomaly \cite{1006.3244}.
}
\end{figure}

Figure~\ref{3p1-fig} shows the allowed regions in the
$\sin^{2}2\vartheta_{e\mu}$--$\Delta{m}^2_{41}$,
$\sin^{2}2\vartheta_{ee}$--$\Delta{m}^2_{41}$ and
$\sin^{2}2\vartheta_{\mu\mu}$--$\Delta{m}^2_{41}$ planes
and the marginal $\Delta\chi^{2}$'s
for
$\Delta{m}^2_{41}$,
$\sin^{2}2\vartheta_{e\mu}$,
$\sin^{2}2\vartheta_{ee}$
and
$\sin^{2}2\vartheta_{\mu\mu}$.

\section{3+2 Neutrino Mixing}
\label{3+2}

In 3+2 neutrino mixing
\cite{hep-ph/0305255,hep-ph/0609177,0906.1997,0705.0107,1007.4171,1103.4570,1107.1452}
the relevant effective oscillation probabilities in short-baseline experiments are given by
\begin{align}
P_{\boss{\nu}{\mu}\to\boss{\nu}{e}}^{\text{SBL}}
=
\null & \null
4 |U_{\mu4}|^2 |U_{e4}|^2 \sin^{2}\phi_{41}
+
4 |U_{\mu5}|^2 |U_{e5}|^2 \sin^{2}\phi_{51}
+
8 \Omega
\sin\phi_{41}
\sin\phi_{51}
\cos(\phi_{54} \stackrel{(+)}{-} \eta)
\,,
\label{trans-3p2}
\\
P_{\boss{\nu}{\alpha}\to\boss{\nu}{\alpha}}^{\text{SBL}}
=
\null & \null
1
-
4 (1 - |U_{\alpha4}|^2 - |U_{\alpha5}|^2)
(|U_{\alpha4}|^2 \sin^{2}\phi_{41} + |U_{\alpha5}|^2 \sin^{2}\phi_{51})
- 4 |U_{\alpha4}|^2 |U_{\alpha5}|^2 \sin^{2}\phi_{54}
\,,
\label{survi-3p2}
\end{align}
for
$\alpha,\beta=e,\mu$,
with
\begin{equation}
\phi_{kj}
=
\Delta{m}^2_{kj} L / 4 E
\,,
\quad
\Omega
=
|U_{\mu4} U_{e4} U_{\mu5} U_{e5}|
\,,
\quad
\eta
=
\text{arg}[U_{e4}^{*}U_{\mu4}U_{e5}U_{\mu5}^{*}]
\,.
\label{3p2}
\end{equation}
Note the change in sign of the contribution of the CP-violating phase $\eta$
going from neutrinos to antineutrinos,
which allows us
to explain the CP-violating difference between MiniBooNE neutrino and antineutrino data.

\begin{figure}[t!]
\begin{center}
\begin{tabular}{cc}
\includegraphics*[width=0.45\linewidth]{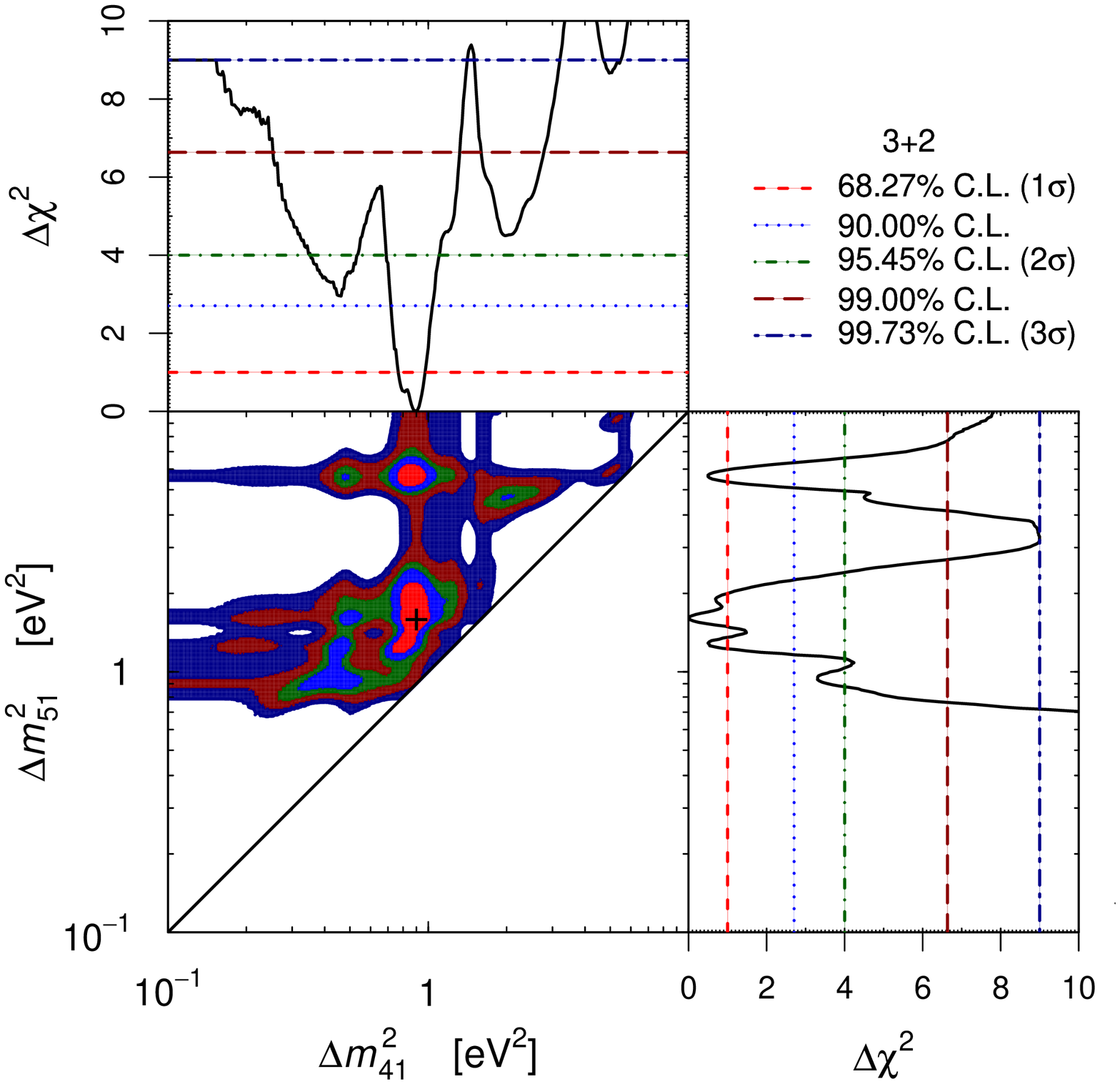}
&
\includegraphics*[width=0.45\linewidth]{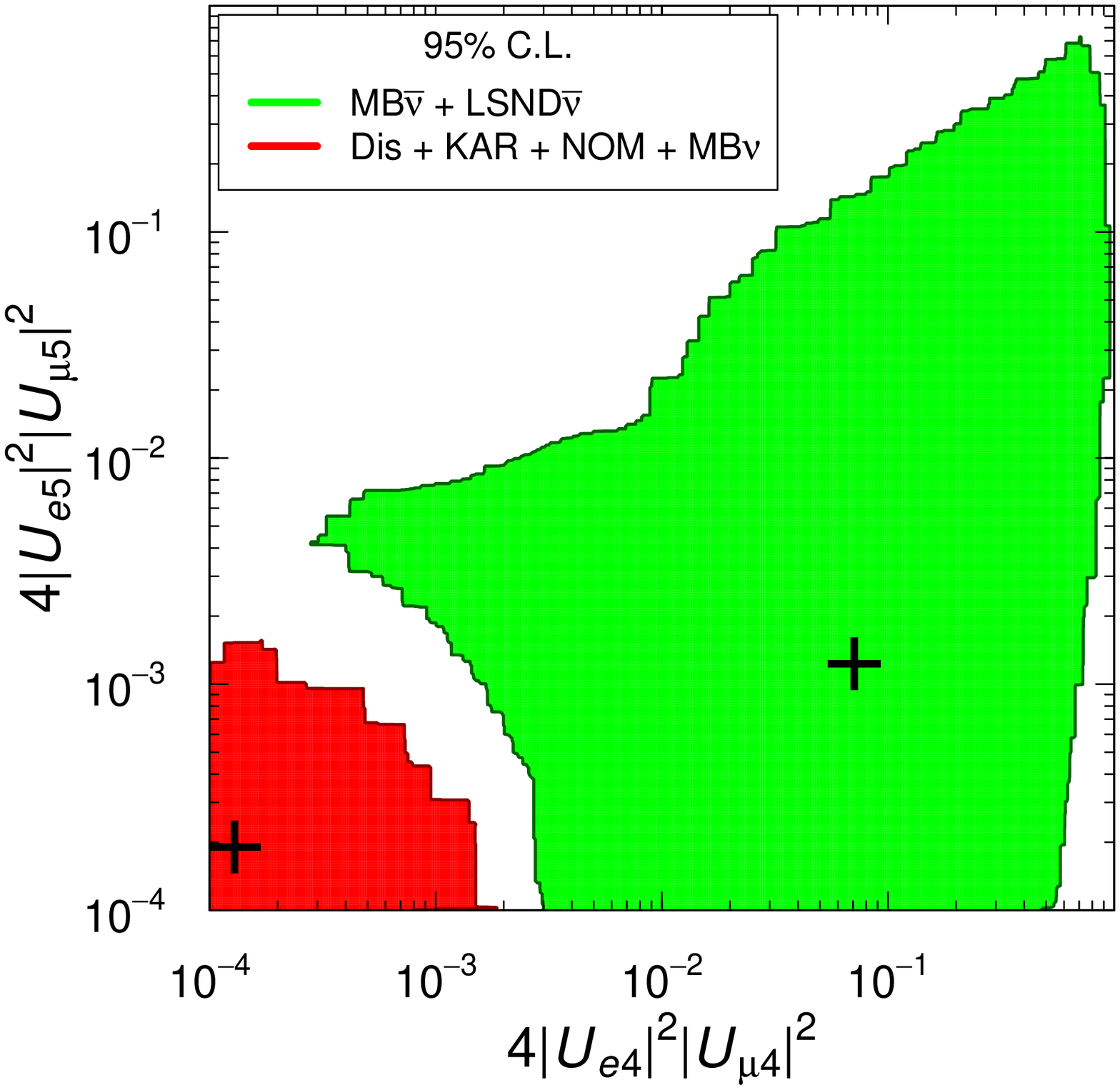}
\end{tabular}
\null
\vspace{-0.5cm}
\null
\end{center}
\caption{\small \label{3p2-fig}
Left panel:
allowed regions in the
$\Delta{m}^2_{41}$--$\Delta{m}^2_{51}$ plane
and
corresponding
marginal $\Delta\chi^{2}$'s
obtained from the global fit of all the considered data in 3+2 mixing.
Right panel:
comparison of the 95\% C.L. allowed regions in the
$4|U_{e4}|^2|U_{\mu4}|^2$--$4|U_{e5}|^2|U_{\mu5}|^2$ plane
obtained from
LSND and MiniBooNE antineutrino data on the right (green area)
and
disappearance,
KARMEN,
NOMAD
and 
MiniBooNE neutrino
data on the left (red area).
Best-fit points are indicated by crosses.
}
\end{figure}

Figure~\ref{3p2-fig}
shows
the marginal allowed allowed regions in the
$\Delta{m}^2_{41}$--$\Delta{m}^2_{51}$ plane
obtained in our 3+2 global fit.
The best-fit values of the mixing parameters are shown in Tab.~\ref{bef}.

The parameter goodness-of-fit obtained with the comparison of the fit of
LSND and MiniBooNE antineutrino data
and the fit of all other data is
$
5
\times
10^{-4}
$.
This is an improvement with respect to the
$
6
\times
10^{-6}
$
parameter goodness-of-fit obtained in 3+1 mixing.
However,
the value of the parameter goodness-of-fit
remains low
and the improvement is mainly due to the increased number of degrees of freedom,
as one can see from Tab.~\ref{bef}.
The persistence of a bad parameter goodness-of-fit
is a consequence of the
fact that the
$\bar\nu_{\mu}\to\bar\nu_{e}$
transitions observed in LSND and MiniBooNE
must correspond in any neutrino mixing scheme to
enough short-baseline disappearance of $\boss{\nu}{e}$ and $\boss{\nu}{\mu}$
which has not been observed
and there is an irreducible tension between the
LSND and MiniBooNE antineutrino data
and the KARMEN antineutrino data.
The only benefit of 3+2 mixing with respect to 3+1 mixing is that
they allow to explain the difference between MiniBooNE neutrino and antineutrino data
through CP violation.
In fact,
neglecting the MiniBooNE neutrino data we obtain
$
\Delta\chi^{2}_{\text{PG}}
=
16.6
$
with
$
\text{PGoF}
=
3
\times
10^{-4}
$
in 3+1 mixing
and
$
\Delta\chi^{2}_{\text{PG}}
=
20.4
$
with
$
\text{PGoF}
=
1
\times
10^{-3}
$
in 3+2 mixing.
In this case
$\Delta\chi^{2}_{\text{PG}}$
is even lower in 3+1 mixing than in 3+2 mixing!

The tension between LSND and MiniBooNE antineutrino data
and
disappearance,
KARMEN,
NOMAD
and
MiniBooNE neutrino
data is illustrated in the right panel of Fig.~\ref{3p2-fig},
which is the analogue for 3+2 mixing of the right panel in Fig.~\ref{exc} in 3+1 mixing.
In practice, in order to show the tension in a two-dimensional figure we have marginalized the $\chi^2$ over all the other mixing parameters,
including the two $\Delta{m}^2$'s.

\section{Conclusions}
\label{Conclusions}

In the framework of 3+1 neutrino mixing,
there is a strong tension between LSND and MiniBooNE antineutrino data
and
disappearance,
KARMEN,
NOMAD
and 
MiniBooNE neutrino
data
\cite{1007.4171,1012.0267,1103.4570,1107.1452}.
Since however the minimum value of the global $\chi^2$ is rather good,
one may choose to consider as possible 3+1 neutrino mixing,
which can partially explain the data,
taking into account its simplicity and the natural correspondence of one new entity
(a sterile neutrino)
with a new effect
(short-baseline oscillations).

In the framework of 3+2 neutrino mixing
the tension between LSND and MiniBooNE antineutrino data
and
disappearance,
KARMEN,
NOMAD
and 
MiniBooNE neutrino
data
is reduced with respect to the 3+1 fit,
but it is not eliminated
(see the right panel of Fig.~\ref{3p2-fig}).
Moreover,
the improvement of the parameter goodness of fit with respect to that obtained in the 3+1 fit
is mainly due to the increase of the number of oscillation parameters,
as one can see from Tab.~\ref{bef}.
Hence it seems mainly a statistical effect.

In conclusion,
I think that the interpretation of the indications in favor of short-baseline oscillations
is uncertain and new experiments are needed in order to clarify the reasons of the tensions in the data
and for leading us to the correct interpretation.

\bibliography{%
bibtex/nu%
}

\end{document}